# Adaptive Multi-band Modulation for Robust and Low-complexity Faster-than-Nyquist Non-Orthogonal FDM IM-DD System

PEIJI SONG[1], ZHOUYI HU[2,*], YIZHAN DAI[1], YUAN LIU[1], CHAO GAO[3], CHUN-KIT CHAN[1]

[1]*Department of Information Engineering, The Chinese University of Hong Kong, Shatin, N. T., Hong Kong, China*
[2]*Department of Electrical Engineering, Eindhoven University of Technology, 5600 MB Eindhoven, The Netherlands*
[3]*Department of Electronic Engineering, The Chinese University of Hong Kong, Shatin, N. T., Hong Kong, China*
*\*z.hu@tue.nl*

**Abstract:** Faster-than-Nyquist non-orthogonal frequency-division multiplexing (FTN-NOFDM) is robust against the steep frequency roll-off by saving signal bandwidth. Among the FTN-NOFDM techniques, the non-orthogonal matrix precoding (NOM-p) based FTN has high compatibility with the conventional orthogonal frequency division multiplexing (OFDM), in terms of the advanced digital signal processing already used in OFDM. In this work, by dividing the single band into multiple sub-bands in the NOM-p-based FTN-NOFDM system, we propose a novel FTN-NOFDM scheme with adaptive multi-band modulation. The proposed scheme assigns different quadrature amplitude modulation (QAM) levels to different sub-bands, effectively utilizing the low-pass-like channel and reducing the complexity. The impacts of sub-band number and bandwidth compression factor on the bit-error-rate (BER) performance and implementation complexity are experimentally analyzed with a 32.23-Gb/s and 20-km intensity modulation-direct detection (IM-DD) optical transmission system. Results show that the proposed scheme with proper sub-band numbers can lower BER and greatly reduce the complexity compared to the conventional single-band way.

## 1. Introduction

Intensity modulation-direct detection (IM-DD) optical transmission is a promising solution for short-reach applications such as data center networks [1,2], visible light communications (VLCs) [3], and passive optical networks [4] due to its low cost. However, the steep frequency roll-off along the spectral response caused by either (i) the interaction between chromatic dispersion (CD) and square-law detection or (ii) the bandwidth limit of transceiver components significantly limits the IM-DD system's transmission performance. The orthogonal frequency division multiplexing (OFDM) technique is a popular solution to combat it in the following two ways: (i) fully making use of the channel characteristics with bit-loading (BL) [5] and (ii) attaining a flat signal-to-noise-ratio (SNR) profile over all subcarriers based on orthogonal matrix precoding/decoding [6-8]. The former is capacity-approaching but requires additional round-trip delay to evaluate the accurate channel state information (CSI), while the latter can improve the performance without requiring CSI and has been extensively studied [6-8]. Recently, by further reducing frequency spacing between neighboring subcarriers beyond the Nyquist limit, the faster-than-Nyquist non-orthogonal frequency-division multiplexing (FTN-NOFDM), also termed spectrally efficient frequency division multiplexing (SEFDM) in literature, was proposed to be more robust against the steep frequency roll-off because of the additional bandwidth saving [9-12]. Though it is possible to combine BL with FTN-NOFDM [13], this scheme is channel-dependent and needs precise CSI, severely restricting the implementation of FTN-NOFDM for time-varying channels. On the contrary, built on the aforementioned orthogonal matrix precoding/decoding, the non-orthogonal matrix precoding

(NOM-p)-based FTN-NOFDM system proposed in Ref. [10,12] is regarded to be advantageous due to its high flexibility and compatibility with the conventional OFDM in advanced digital signal processing (DSP) already used in OFDM, such as zero-forcing (ZF)-based channel equalization. However, only the single-band case was investigated in previous works, leading to the underutilization of the NOM-p-based FTN-NOFDM to maximize the transmission performance.

In this work, by dividing the single band into multiple sub-bands in the NOM-p-based FTN-NOFDM system, we propose an adaptive multi-band modulation FTN-NOFDM technique, where different quadrature amplitude modulation (QAM) levels are applied to different sub-bands, effectively utilizing the low-pass-like channel and reducing the complexity. We experimentally study the effect of sub-band number on bit-error-rate (BER) performance and implementation complexity in a 32.23-Gb/s and 20-km intensity modulation-direct detection (IM-DD) optical transmission system. Experimental results show that the proposed scheme using appropriate sub-band numbers can lower the BER and reduce the decoding complexity compared to the preceding single-band scheme. Furthermore, we verify that the flatter the BER over sub-bands, the better the achieved performance, while the sub-band number has marginal influence on the complexity in the case that the sub-band number is greater than or equal to 2.

## 2. Principles

### 2.1 Adaptive multi-band modulation FTN-NOFDM technique

The baseband signal of the conventional FTN-NOFDM signal is given by [10]

$$x(t) = \sum_{v=0}^{V-1} S_v \exp(\frac{i2\pi v\alpha t}{T}), \quad (1)$$

where $S_v$ is the symbol on the $v$-th subcarrier, $V$ is the symbol size, $\alpha = \Delta f \cdot T$ is the bandwidth compression factor, $\Delta f$ is the adjacent sub-carrier's spacing, and $T$ is the symbol period. $\alpha=1$ corresponds to the conventional OFDM signal, and it is the FTN-NOFDM for $\alpha<1$. Sample $L$ times within a symbol period $T$ of $x(t)$, Eq. (1) is rewritten as

$$X[k] = \sum_{v=0}^{V-1} S_v \exp(\frac{i2\pi v\alpha k}{L}), \quad k=0, 1, ..., L-1, \quad (2)$$

where $X[k]=x(kT/L)$. The mature OFDM modulator can be employed to effectively generate the FTN-NOFDM signal by transforming Eq. (2) to

$$\hat{X}[k] = \sum_{v=0}^{N_{fft}-1} \hat{S}_v \exp(\frac{i2\pi vk}{N_{fft}}), \quad k=0, 1, ..., L-1, \quad (3)$$

where $N_{fft} \geq (floor(L/\alpha)+1)$, $\hat{S}_v = S_v$ for $v \leq V-1$, otherwise $\hat{S}_v = 0$, and $X[k] = \hat{X}[k]$ for $k \leq L-1$. Fig. 1(a) depicts the principle of real-valued FTN-NOFDM signal generation based on an OFDM modulator. Due to the loss of subcarrier orthogonality caused by the discarded $N_{fft}-L$ symbols at the transmitter side, channel estimation and equalization will become a crucial challenge. Though some solutions have been proposed [14,15], they come at the expense of implementation complexity.

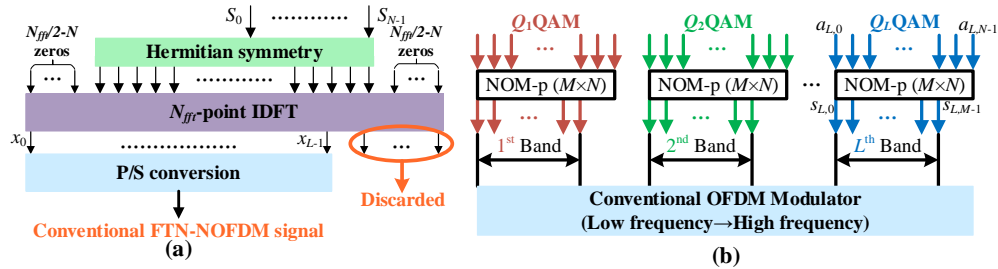

Fig. 1. Principle of (a) the conventional FTN-NOFDM, (b) the proposed adaptive multi-band modulation based FTN-NOFDM, where $L=1$ corresponds to the conventional single-band case.

To make the channel equalization of FTN-NOFDM compatible with the conventional OFDM modulator, NOM-p was proposed by shifting the compression factor $α$ into $S_v$ for $v=0$, …, $V−1$ to entirely use the mature OFDM modulator. Fig. 1(b) shows the generation of FTN-NOFDM signal utilizing the NOM-p, where $L=1$ is the conventional single-band scheme, and $L≥2$ corresponds to the proposed adaptive multi-band modulation by evenly partitioning the single band into multiple sub-bands.

Since the single band is a particular case of multi-band when $L=1$, the $L$-th sub-band in Fig. 1(b) is used as an example to illustrate the concept of NOM-p. The $N$ subcarriers (e.g., $S_{L,0}$, …, $S_{L,N−1}$ for the $L$-th sub-band) are squeezed into $M$ subcarriers (e.g., $a_{L,0}$,…, $a_{L,M−1}$ for the $L$-th sub-band) by using an $M×N$ non-orthogonal matrix (NOM), where $N=V/L$. The bandwidth compression factor $α$ in this process is defined by $α=M/N$, corresponding to $(1−α)×100\%$ of the bandwidth saved. The NOM is generated by an $N×N$ orthogonal matrix with $N−M$ rows or columns discarded, which could be a standard discrete Fourier transform (DFT) matrix, an orthogonal circulant transform (OCT) matrix, or other suitable orthogonal precoding matrices [6-8]. Without loss of generality, we use the $N×N$ OCT matrix [16] in this work. After the operation of NOM-p for all $L$ sub-bands, the obtained $L·M$ subcarriers are allocated onto the orthogonal subcarriers from low to high frequency in the OFDM modulator as

$$X[k] = \sum_{b=0}^{B-1} a_b \exp(\frac{i2\pi bk}{\hat{N}_{fft}}),\ k = 0,\ 1,\ ...,\ \hat{N}_{fft} − 1, \quad (4)$$

where $B=M·L$, $\hat{N}_{fft}$ is the suitable FFT size in the OFDM modulator. Hence, the NOM-p-based FTN-NOFDM signal shows a similar behavior as the conventional OFDM signal, facilitating its channel estimation/equalization. It should be noted that some advanced DSP techniques already used in OFDM can also be used in this FTN-NOFDM system.

Considering that an IM-DD channel usually has a low-pass-like frequency response due to CD and the bandwidth limitation of transceiver components, the QAM levels $Q_l$ for $l=1$, …, $L$ assigned to per sub-band should satisfy $Q_1 ≥ Q_2 ≥ \cdots ≥ Q_L$ to make more rational use of the channel. With the help of bandwidth saving, the high-frequency region, usually with the lowest SNR, could be discarded. Therefore, the robustness against the steep frequency roll-off can be further improved in the proposed adaptive multi-band FTN-NOFDM system.

*2.2 Detection of FTN-NOFDM signal*

The inter-carrier interference (ICI) is produced from the NOM-p when $M<N$. The logarithmic-maximum-a-posterior (log-MAP) Viterbi decoding is a common algorithm to reduce the ICI because of its maximum likelihood (ML) performance; see Section 2.2 of Ref. [11] for detailed principles. The log-MAP Viterbi algorithm requires precise information on the noise variance of each subcarrier ahead of the inverse NOM-p at the receiver side. Due to the noise-spreading effect of NOM, the noise variance per subcarrier within the same sub-band is set to be the same, which is calculated as

$$\hat{\sigma}_{l,n} = \frac{1}{N} \sum_{m=0}^{M-1} \sigma_{l,m}, l = 1,\ ...,\ L,\ n = 0,\ ...,\ N−1, \quad (5)$$

where $\sigma_{l,m}$ is the estimated noise variance on the $m$-th subcarrier of the $l$-th sub-band prior to the inverse NOM-p. The surviving path in the log-MAP Viterbi algorithm for each sub-band is set equal to the corresponding QAM level in the following experiment, as it can already guarantee the saturated performance [11].

*2.3 Complexity analysis*

**Table 1. The Approximately Computational Complexity**

|    | NOM-p   | Inverse NOM-p | Log-MAP Viterbi |
|----|---------|---------------|-----------------|
| CM | $MN$    | $NM$          | $((-30-90CQ)+(59+25CQ)N+(45CQ-30)N^2+20CQN^3+N^5)/30$ |
| CA | $M(N-1)$ | $N(M-1)$     | $((-30-30CQ)+(59-20CQ)N+(30CQ-30)N^2+20CQN^3+N^5)/30$ |

The computational complexity of one sub-band containing NOM-p, inverse NOM-p, and log-MAP Viterbi algorithm, in terms of complex multiplication (CM) and complex addition (CA), is summarized in Table 1, where $C$ is the number of surviving paths in Viterbi decoding and $Q$ is the applied QAM level. For the log-MAP Viterbi algorithm, we only consider the complexity induced by Eq. (7)-(9) in Ref. [11] due to their contribution to the main complexity.

Since the complexity of log-MAP Viterbi is much higher than NOM-p and inverse NOM-p, and the exponential terms 3 and 5 in log-MAP Viterbi make them much bigger than the other terms, the implementation complexity can be approximated as

$$\begin{aligned} CM &= (20CQN^3 + N^5)/30, \\ CA &= (20CQN^3 + N^5)/30. \end{aligned} \quad (6)$$

## 3. Experimental setup

Fig. 2 shows the experimental setup and the corresponding DSP. The offline-generated digital signal using MATLAB was first transformed into the analog signal by an AWG, whose Vpp is optimized at 300 mV during this experiment. Subsequently, the output signal was amplified by an EA. One MZM combined with an ECL at 1550nm converted the electrical signal to the optical domain. After 20-km SSMF transmission, the signal was detected by a receiver module, which consists of a PD and an EA. Finally, RTO recorded the signal current which is processed by offline DSP. Inset (i) on the right side of Fig. 2 gives the channel frequency response, where the SNR reduces quickly when the frequency exceeds 10 GHz. Fig. 3 shows the electrical spectra of the received FTN-NOFDM signal with different bandwidth compression factors. It can be seen from Fig. 3 that the signal bandwidth is compressed from 12.18 GHz to 10.97GHz, and 9.75 GHz when $α$ is set to 0.9, and 0.8, respectively.

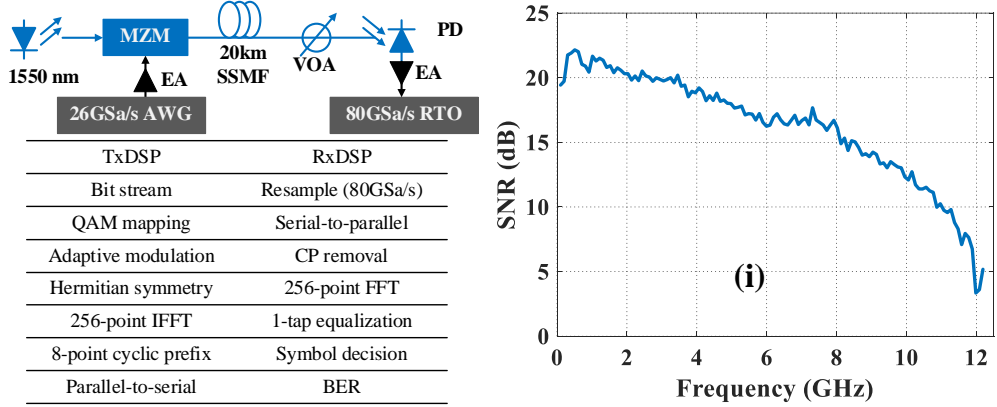

Fig. 2. Experimental setup and DSP flow chart. (AWG: arbitrary waveform generator, EA: electrical amplifier, MZM: Mach-Zehnder modulator, SSMF: standard single-mode fiber, VOA: variable optical attenuator, PD: photodiode, RTO: real-time oscilloscope). Inset (i): measured channel response after 20-km transmission.

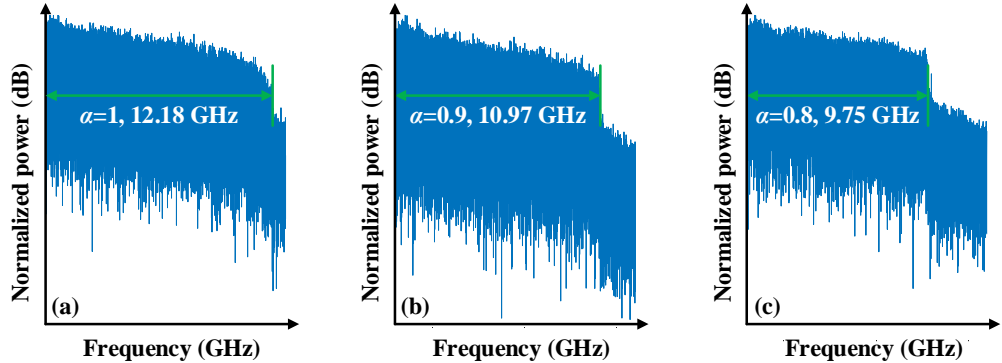

Fig. 3. The electrical spectra of the received FTN-NOFDM signals with a compression factor of (a) 1, (b) 0.9, and (c) 0.8, respectively.

For a fair comparison, the original total number of subcarriers ($V$) was fixed at 120. The IFFT size was set to 256, and the cyclic prefix (CP) length was 8. 20 blocks of 4QAM OFDM training symbols (TS) were added before 200 blocks of payload symbols (spectral efficiency=3bits/symbol) for equalization. The DAC worked at 26GSa/s, and the RTO worked at 80GSa/s. Therefore, the data rate of the signal excluding CP and TS was 32.23 ($≈26×3×120/(256+8)×200/220$) Gb/s.

For the single-band case, the original subcarriers were modulated with an 8QAM format, where the circular constellation was used due to the best performance in both OFDM and FTN-NOFDM systems [17]. Considering the low-pass property of the IM-DD channel, bit allocations for the multiple sub-band scenarios follow a decreasing trend for different sub-bands. As a result, to ensure the same data rate, we set $[Q_1, Q_2]$=[16, 4] for the 2-band case, $[Q_1, Q_2, Q_3]$=[16, 8, 4] for the 3-band case, $[Q_1, Q_2, Q_3, Q_4]$=[16, 16, 4, 4] or [16, 16, 8, 2] for the 4-band case, and $[Q_1, Q_2, Q_3, Q_4, Q_5]$=[16, 16, 8, 4, 4] for the 5-band case, respectively.

## 4. Experimental results

Fig. 4 shows the experimental results. We first studied the BER performance of the proposed adaptive multi-band modulation with different $L$ and $α$ values, as shown in Fig. 4(a), where only some sub-band numbers are chosen ($L∈\{1, 2, 3, 4, 5\}$) to verify the effectiveness of the proposed scheme. The digital numbers inside square brackets represent the QAM levels allocated per sub-band. It can be found from Fig. 4(a) that there is an optimum $α$ value for different sub-band number cases. Because the system is bandwidth-limited, appropriate bandwidth saving can help alleviate the negative effect of the steep frequency roll-off, but when $α$ further decreases, the penalty of the increased ICI becomes dominant, reducing the overall BER performance. Based on the results in Fig. 4(a), we compared the lowest achievable BER of different sub-band numbers $L$ in Fig. 4(b). We can see that except for $L$=4, the other multi-band signals can outperform the conventional single-band NOM-p-based FTN signal ($L$=1); among them, $L$=3 performs the best in the chosen examples. Based on the complexity analysis in Eq. (6), the reduced complexity applying the proposed multi-band scheme with $L$=3 in complex multiplication and complex addition are 97.28% and 97.28%, respectively. It should be noted that the reduced complexity for $L$=2 is 89.92% close to 97.28%, which means the sub-band number has a marginal influence on the complexity when it is greater than or equal to 2.

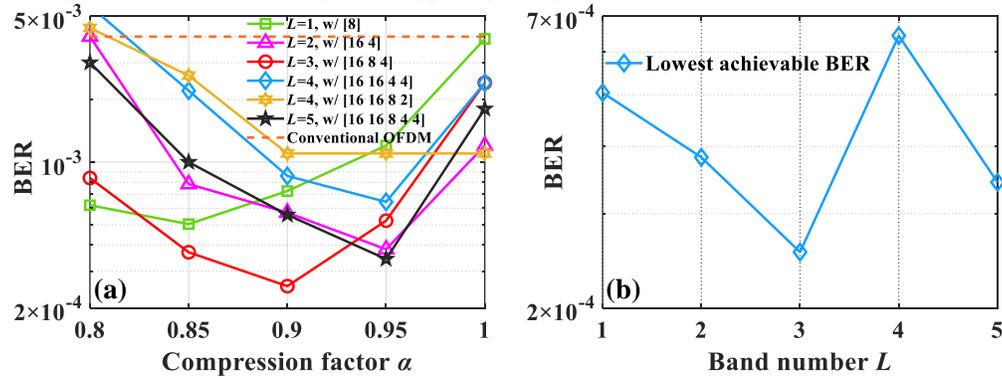

Fig. 4. (a) The BER versus the number of sub-bands ($L$) with different compression factors ($α$), and(b) the lowest achievable BER under different $L$ values.

Table 2. Expanded BER per sub-band from Fig. 4(b)

| Band Number \ Band Index | 1 | 2 | 3 | 4 | 5 |
|---|---|---|---|---|---|
| 1 | $4.73\times10^{-6}$ | | | | |
| 2 | $4.6\times10^{-4}$ | $2.25\times10^{-4}$ | | | |
| 3 | $2.6\times10^{-4}$ | $1.75\times10^{-4}$ | $3.62\times10^{-4}$ | | |
| 4 | $1\times10^{-4}$ | $1.4\times10^{-4}$ | 0 | $8\times10^{-4}$ | |
| 5 | $1\times10^{-4}$ | $6\times10^{-4}$ | $2\times10^{-4}$ | 0 | $1\times10^{-3}$ |

To further investigate why $L$=3 leads to better performance, Table 2 expands the BER of each subband from Fig. 4(b). We can see from Table 2 that $L$=3 can achieve a flatter BER than other cases, avoiding the impact of a relatively large BER on the system performance. This inspires us that optimizing a suitable sub-band number and corresponding QAM levels leading to a flatter BER is vital to improve performance.

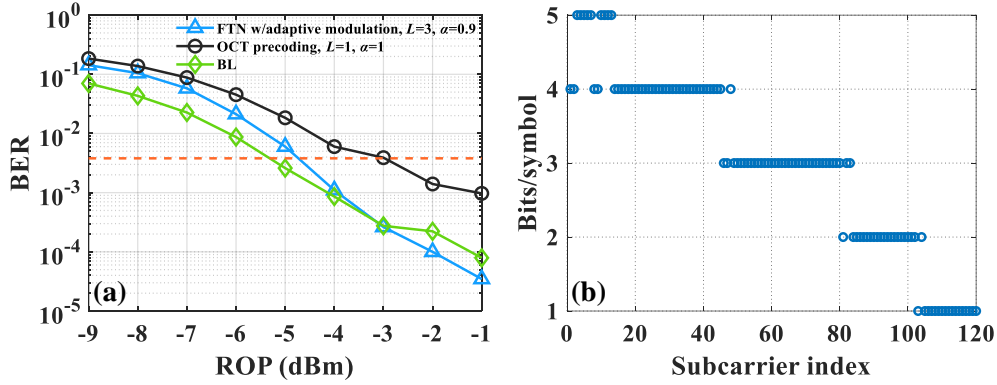

Fig. 5. (a) Measured ROP sensitivity curves ($L$=3 for adaptive multi-band modulation), and (b) bits allocation result with the adaptive-loaded DMT at ROP=−1dBm.

Then, we measured the received optical power (ROP) sensitivity curves as presented in Fig. 5(a), where $L$ is set to 3 for the proposed scheme. Fig. 5(b) shows the transmitted QAM constellations of each sub-band for $L$=3 where the ROP=−1 dBm. Besides, we also measured the BER performance of BL with Chow's algorithm to provide a benchmark, whose bit allocation result at ROP=−1 dBm is shown in Fig. 5(c). It can be found that the proposed method shows performance similar to the BL scheme due to the saving of 10% bandwidth located at high frequency (worst SNR). Meanwhile, compared with the conventional OCT-precoded OFDM [6,7,16], the proposed adaptive multi-band FTN scheme can significantly improve the ROP sensitivity by ~1.7 dB, indicating its superiority.

## 5. Conclusions

This paper proposes an adaptive multi-band modulation FTN scheme to make more rational use of the low-pass-like channel and reduce the computational complexity. The results of a 32.23-Gb/s and 20-km IM-DD transmission experiment show that the proposed scheme outperforms the conventional single-band FTN scheme while significantly reducing the implementation complexity. Especially at $L$=3, the BER can be reduced from $5.04\times10^{-4}$ to $2.54\times10^{-4}$ at ROP=−3dBm, and the corresponding complexity reduction is 97.28% in complex multiplication and 97.28% in complex addition, respectively. Due to the bandwidth saved in high frequency (worst SNR), the adaptive multi-band modulation FTN scheme performs comparably with the BL scheme. Although verified in the IM-DD system, the proposed scheme is feasible for other systems suffering from the steep frequency roll-off, for instance, the VLC system and underwater transmission system.